\documentstyle[12pt,epsfig]{article}
\def\eqn{\begin{equation}}
\def\enn#1{\label{#1} \end{equation}}
\def\eq{\[}
\def\en{\]}
\def\r#1{(\ref{#1})}

\def\f#1{\ref{#1}}

\def\vs#1{\vspace{#1}}

\begin{document} {
\setlength{\baselineskip}{12pt}

\pagestyle{plain}
\title{Reverse Integration for Computing Stationary Points of Unstable Stiff Systems}

\author{C. W. Gear\thanks{NEC Research Institute, retired}  and
Ioannis Kevrekidis\thanks{Department of Chemical Engineering, PACM and Mathematics, Princeton University}}

\maketitle

\begin{abstract}
Using existing, forward-in-time integration schemes, we demonstrate
that it is possible to compute unstable, saddle-type fixed points of
stiff systems of ODEs when the stable compenents are fast (i.e.,
rapidly damped) while the unstable components are slow.  The 
approach has implications for the reverse (backward in time)
integration of such stiff systems, and for the ``coarse reverse"
integration of microscopic/stochastic simulations. 
 
\end{abstract}


\section{Introduction}

We consider the problem of computing a stationary point, $y_0: f(y_0)=0$, of the
differential equation
\eq
\frac{dy}{dt} = f(y)
\en
when $y_0$ is unstable.  Here $f$ is $\Re^n \mapsto \Re^n$. 

If the differential equation is exponentially stable to a stationary
point in reverse time 
then an integration backwards in time will approach that
stationary point.  
In a neighborhood of such a point the eigenvalues of
\eq
J = \frac{\partial f}{\partial y}
\en
will be in the positive half plane.  
In that case we can simply numerically integrate in the reverse
direction (negative t), and, if we start sufficiently close to the stationary
point and use a small enough step size, we will asymptotically approach it.  

We are interested in various situations in which backward integration
is not feasible.  
If all we have is a ``black box''
time-stepper, appropriate only for integration in the forward direction (this could be
a legacy code, or a stochastic kinetic Monte Carlo simulator which only works in forward
time) we cannot integrate backwards directly.
If there are eigenvalues in both half planes, then the differential equation is 
unstable in either direction and
an accurate integration will have the same properties.  
We are
particularly interested in the case that the instability in the
forward direction is ``mild'' - that is, the eigenvalues in the
positive half plane are fairly close to the origin, while the
instability in the reverse direction is ``severe'' - that is, the
remaining 
eigenvalues have quite negative real parts and the 
corresponding solution components are
{\em strongly stable} in the forward direction.  
This type of situation
arises in many physical problems where strongly damped
components are coupled with mildly growing components - that is, in stiff
system where the slow components are unstable.   
This type of problem
could arise, for example, in a singular perturbation context, or when a
Differential-Algebraic equation is regularized by converting the
algebraic terms into stiffly stable differential components. 

In an earlier paper (\cite{cwgygk1} we considered
projective methods for stiff problems with gaps in their spectra.  In
the projective method, a numerical solution is computed at a sequence
of relatively closely spaced points in time using a conventional
integrator with small time steps, and then a ``giant''
step is taken using polynomial extrapolation from the last few of
the points computed by the small steps.  
This giant step was called the {\it projective
step}.  It was taken forward in time.  
The small steps stabilized the fast
(``strongly stable'') components; the large, projective step had a
stability region associated with explicit, large step methods that
are stable for slowly damped components. 
The 
combined method had a stability region for linear problems that had two
components, one that caused damping of the fast components and one
that caused damping of the slow.

In the discussion below we will continue to talk about the ``rapidly
damped'' components, referring to those that are damped in the {\em
forward time direction}, and to unstable components as those that grow
in the {\em forward time direction}.  
We will do this even when we
discuss integrating in the reverse direction, so as to avoid always
having to qualify the terms ``stable components'' and ``unstable components.''

In this note we consider using
projective methods in which the projective step is taken {\em backwards in
time} while the small steps using the conventional integrator remain
forward in time.  
Thus the overall reverse projective integration step
consists of small regular
forward steps followed by a ``giant'' reverse step, giving a net
integration backwards in time.  
We will show that the resulting
stability regions consist, once again, of two components.  The first
is due to the conventional small step forward integrator and as before
leads to stability of the rapidly damped components.  The second, due
to the projective step, now leads to stability of the slowly growing, unstable
components of the differential equation. 

\section{Analysis}

We will analyze the simplest of these methods following the technique
used in \cite{cwgygk1}.
The reverse projective integration step we will discuss here consists
of $k+1$ {\em inner} integration steps of size $h$ forward from $t_n$
to $t_{n+k+1}$, followed 
by one projective step of size $-Mh$ to arrive at $t_{k+1-M}$.
The projective step takes the form
\eq
y^p_{k+1-M} = My_k - (M-1)y_{k+1}
\en

As in \cite{cwgygk1} we assume that the error amplification of an
inner integrator step is $\rho(h\lambda)$.  Assuming that the inner
integrator is first order accurate we have
\eq
\rho(h\lambda) = 1 + h\lambda + {\rm O}(h\lambda)^2
\en
and if the inner integrator is the forward Euler method, the last term
can be dropped.
Then we see that the reverse projective method has an error
amplification $\sigma$ given by
\eq
\sigma = \rho^k(M - (M-1)\rho).
\en
As before, we define the stability region in the $\rho$-plane as the
set of $\rho$ for which $\sigma$ is not outside the unit disk, and plot its
boundary by finding the set of $\rho$ such that $|\sigma| = 1$.
Figure \f{f1} shows the plots for $k = 2$ and four different values
of $M$.  
Note that, because the process goes forward $k+1 = 3$ steps
before going backwards $M$ steps, the values actually correspond to
net reverse steps of $h, 2h, 4h$, and $8h$ respectively.  
The
method is stable inside the regions shown.   As $M$ gets large, these
regions asymptotically tend to a pair of disks.  One, centered at
$1+1/M$ and of radius 
$1/M$, corresponds to the stability region of the forward Euler
method because the reverse projective step is the equivalent of a
forward Euler method in the reverse direction.  The second is centered at
the origin and has radius $M^{-1/k}$.  It represents the region where the
damping of $\rho^k$ is sufficient to overcome the growth proportional
to $M$.  In other words, the regions are essentially similar to those for forward
projective methods, except that the stability region due to the
projective step corresponds to $h\lambda$ in the {\it positive} half plane
since that step is taken in the reverse direction.

\begin{figure}[p]
\centerline{\psfig{figure=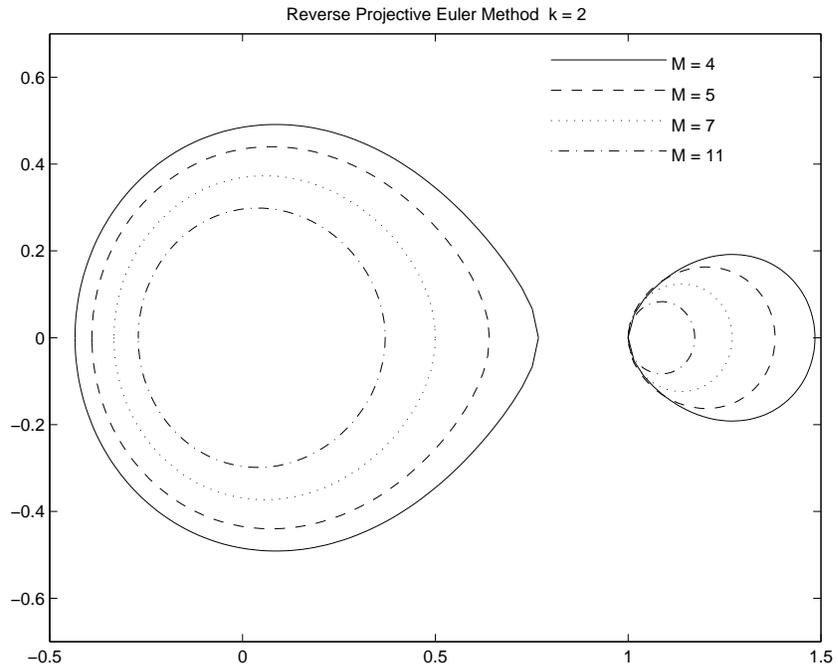,height=3.5in,height=3.5in}}
\caption{Stability Regions for Reverse Projective Methods.}\label{f1}
\end{figure}

\section{Stationary Methods}
In the above we have taken $M > k+1$ so that the net progress is
negative in time.  If we choose $M = k+1$ the total step length of the
reverse projective integration method will be zero; we will call this
a {\em stationary projective method}.  
It has the curious stability
region shown in Figure \f{f2}.  It is stable along a section of the
real axis that includes $\rho = 1$ (which is $h\lambda = 0$).  The
boundary crosses itself at $\rho = 1$.  The
intersection is at right angles because in the neighborhood of $\rho
= 1$ simple algebra shows that $\sigma$ is given by
\eqn
\sigma = 1 - k(k+1)/2(\rho - 1)^2 - k^2(k-1)/2(\rho - 1)^3
\enn{order}
so the stability in the neighborhood of $\rho  = 1$ is equivalent to
requiring that
\eq
{\rm Real}((\rho - 1)^2) \ge 0
\en
which happens when ${\rm Arg}(\rho) \in [-\pi/4, \pi/4] $ or
$[3\pi/4, 5\pi/4 ]$.  
The form of eq. \r{order} arises because the
first-order accuracy of the method for a net step size of length zero
means that
\eq
\sigma(0\lambda) = 1 + 0\lambda + {\rm O}(h\lambda)^2 =1+ {\rm O}(\rho -
1)^2
\en
\begin{figure}[p]
\centerline{\psfig{figure=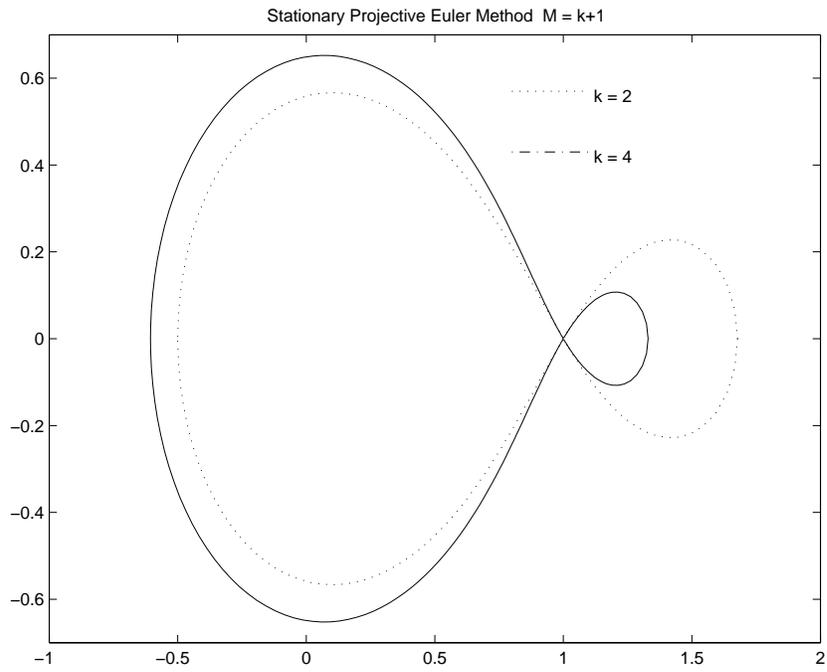,height=3.5in,height=3.5in}}
\caption{Stability Regions for Stationary Projective Methods.}\label{f2}
\end{figure}

It is possible that this procedure will be of value in the context of
finding consistent initial conditions for differential algebraic equations (see
\cite{bhp}).  If the algebraic components of a DAE were converted to
stiff components (as has been advocated by some), the initialization
problem is to 
find an initial condition that is ``on the slow manifold'' (i.e., the
solution manifold of the original DAE.)

\section{Comments}

The procedure we just outlined will converge to saddle-type points
whose linearization is characterized by a gap between strong stable
modes and weak unstable ones. An existing forward simulation code will
not in general converge to such points (the numerical trajectories
will move away forward in time, and ``explode away" backward in
time). We can therefore think of our procedure as a computational
superstructure that transforms a forward simulation code into a
contraction mapping capable of converging to such saddle unstable points. The
zero-net backward movement process described at the end can be related
to consistent initialization algorithms for differential-algebraic
equations.  

Beyond the computation of saddle-type fixed points, however, if the
forward in time dynamics possess such a separation of time scales
globally, and an attracting, forward-invariant, low-dimensional slow
manifold exists, the procedure becomes an ``on manifold" reverse time
integration scheme. 
That is, the regularizing action of the forward
integration allows us to follow the on-manifold well-behaved
trajectories backward in time. 
This may then provide a meaningful -and
very simple to implement- way of regularizing the reverse, ``on the
slow manifold", dynamics of stiff sets of ODEs and even
discretizations of disspative PDEs.  
For example, in contexts where a
low-dimensional inertial manifold exists for a dissipative PDE \cite{Temam}, our
superstructure enables a direct simulation of an accurate
discretization of the PDE to follow backward trajectories on the
inertial manifold without ever having to explicitly derive an inertial
form (or approximate inertial form).  

In the spirit of the last observation, it is interesting to consider
the implications of such a process for a meaningful reverse
integration of systems described by microscopic/stochastic
simulators. 
In many practically relevant cases, the coarse-grained
behavior of such simulators can be described by the evolution in time
of a few ``master" moments of microscopically evolving
distributions. 
The remaining, higher moments, become quickly slaved by
forward simulation to the slow, master moments. 
We have recently
proposed coarse projective integration schemes that use short bursts
of appropriately initialized microscopic/stochastic simulation to
estimate the time-derivative of the unavailable coarse equations for
the master modes, and ``project" these modes forward in time \cite{GKT,SGK,HK}
If the
coarse projection is performed backward in time, the procedure will
allow us to follow the regularized reverse time behavior of the coarse
variables. 
This is done using the microscopic/stochastic
forward-in-time simulator directly, circumventing the necessity of
deriving an explicit macroscopic closure. 
It becomes therefore
possible to use a forward-in-time molecular dynamics simulator to
extract regularized reverse-time information of coarse system
variables. 
We have already demonstrated the feasibility of this
technique in the case of coarse molecular dynamics simulations of a
dipeptide folding kinetics in water \cite{HK}. 
Coarse reverse
integration allows us to use microscopic simulators to quickly escape
free-energy minima, to converge on certain transition states (saddles
on the free-energy surface) and, more generally, may enhance our 
ability to explore the structure of free energy surfaces. 

In summary, the technique holds promise towards (a) the computation of
unstable, saddle-type fixed points using existing simulators; (b) the
regularized, ``on manifold" backward-in-time integration of certain dissipative PDEs
possessing low-dimensional, exponentially attracting slow manifolds;
and (c) the use of microscopic/stochastic simulators to track
coarse-grained behavior backward in time, enhancing the ability to
escape free energy minima and to locate saddle-type coarse-grained
``transition states".  

\vs{0.2in}
{\bf Acknolwedgements.} 
This work was partially supported by AFOSR
(Dynamics and Control, Dr. B. King) and an NSF ITR grant. Discussions
with Profs. P. G. Kevrekidis (UMass), Ju Li (OSU) and Dr. G. Hummer (NIH)
are gratefully acknowledged.


\newpage   

\begin{thebibliography}{99}

\bibitem{bhp} Brown P. N., Hindmarsh A. C. and Petzold 
L. R. (1998). Consistent initial condition calculation for
differential-algebraic systems, SIAM J. Sci. Comput., 19, 1495

\bibitem{cwgygk1} Gear, C. W. and Kevrekidis, I. G,
Projective Methods for Stiff Differential Equations: {\em problems
with gaps in their eigenvalue spectrum}, NEC Research Institute Report
2001-029, in press, {\it SIAM J. Sci. Comp.}

\bibitem{cwgygk2} Gear, C. W. and Kevrekidis, I. G, Telescopic
Projective Methods for Stiff Differential Equations, NEC Research
Institute Report 2001-122, in press, {\it J. Comp. Phys.}

\bibitem{Temam} Temam, R. (1988) {\it Infinite Dimensional Dynamical Systems
in Mechanics and Physics} Springer Verlag, NY.

\bibitem{GKT} Gear, C. W., Kevrekidis, I. G. and Theodoropoulos, K. (2002)
Coarse Integration/Bifurcation Analysis via Microscopic
Simulators: micro-Galerkin methods,  {\it Comp. Chem. Engng.} {\bf 26} pp.941-963

\bibitem{SGK} Siettos, C. I., Graham, M. D. and Kevrekidis, I. G. 
Coarse Brownian Dynamics Computations for Nematic Liquid Crystals"
submitted to {\it J. Chem. Phys.};
can be obtained as cond-mat/0211455 at arXiv.org.

\bibitem{HK} Hummer, G. and Kevrekidis, I. G. 
Coarse molecular dynamics of a peptide fragment: free energy, kinetics
and long time dynamics computations", submitted to {\it J. Chem. Phys.}, 
Can be obtained as physics/0212108 at arXiv.org.



\end{thebibliography}
\end{document}